\documentclass[conference]{IEEEtran}
\IEEEoverridecommandlockouts
\usepackage{algorithm}
\usepackage{algpseudocode}
  
\usepackage{cite}
\usepackage{amsmath,amssymb,amsfonts}
\usepackage{subfigure}
\usepackage{graphicx}
\usepackage{soul}
\usepackage{textcomp}
\usepackage{xcolor}
\def\BibTeX{{\rm B\kern-.05em{\sc i\kern-.025em b}\kern-.08em
    T\kern-.1667em\lower.7ex\hbox{E}\kern-.125emX}}

\begin{document}

\title{Dynamic Beamforming and Power Allocation in ISAC via Deep Reinforcement Learning\\
}

\author{\IEEEauthorblockN{1\textsuperscript{st} Duc Nguyen Dao}
\IEEEauthorblockA{\textit{Radio Systems Group} \\
\textit{University of Twente}\\
Enschede, Netherlands \\
d.d.n.dao@utwente.nl}
\and

\IEEEauthorblockN{2\textsuperscript{nd} André B. J. Kokkeler}
\IEEEauthorblockA{\textit{Radio Systems Group} \\
\textit{University of Twente}\\
Enschede, Netherlands  \\
a.b.j.kokkeler@utwente.nl}
\and

\IEEEauthorblockN{3\textsuperscript{rd} Haibin Zhang}
\IEEEauthorblockA{\textit{Department of Networks} \\
\textit{TNO}\\
The Hague, Netherlands  \\
haibin.zhang@tno.nl}

\and
\IEEEauthorblockN{4\textsuperscript{th}  Yang Miao}
\IEEEauthorblockA{\textit{Radio Systems Group} \\
\textit{University of Twente}\\
Enschede, Netherlands  \\
y.miao@utwente.nl}

}
\maketitle

\begin{abstract}
Integrated Sensing and Communication (ISAC) is a key enabler in 6G networks, where sensing and communication capabilities are designed to complement and enhance each other. One of the main challenges in ISAC lies in resource allocation, which becomes computationally demanding in dynamic environments requiring real-time adaptation. In this paper, we propose a Deep Reinforcement Learning (DRL)-based approach for dynamic beamforming and power allocation in ISAC systems. The DRL agent interacts with the environment and learns optimal strategies through trial and error, guided by predefined rewards. Simulation results show that the DRL-based solution converges within 2000 episodes and achieves up to 80\% of the spectral efficiency of a semidefinite relaxation (SDR) benchmark. More importantly, it offers a significant improvement in runtime performance, achieving decision times of around 20 ms compared to 4500 ms for the SDR method. Furthermore, compared with a Deep Q-Network (DQN) benchmark employing discrete beamforming, the proposed approach achieves approximately 30\% higher sum-rate with comparable runtime. These results highlight the potential of DRL for enabling real-time, high-performance ISAC in dynamic scenarios.
\end{abstract}


\section{Introduction}

In 6G networks, Integrated Sensing and Communication (ISAC) has progressed from a promising research concept to a core technology, with standardization efforts now underway by organizations such as 3GPP and ITU \cite{isac_survey}. Resource allocation is one of the most challenging aspects of ISAC system design, as it requires the joint optimization of communication and sensing functions. This process often involves significant complexity and trade-offs, due to the shared use of time, frequency, spatial, and hardware resources.

Multiple-input multiple-output (MIMO) technology plays a vital role in ISAC systems by enabling simultaneous support for communication and sensing through its spatial diversity and multi-beamforming capabilities. By leveraging MIMO, ISAC systems can achieve higher array gain, improved signal directivity, and effective interference mitigation—critical factors for enhancing the performance of both communication links and sensing accuracy.

\subsection{Related works}
Many studies in the literature approach beamforming as an optimization problem, enabling the incorporation of diverse objectives and constraints for greater design flexibility. In \cite{EE_optimization}, the authors formulated an energy-efficient beamforming maximization problem for ISAC systems and addressed it using the successive convex approximation (SCA) method. Moreover, a beam pattern matching problem (for sensing), which aims to simultaneously maximize the desired communication sum-rate, was proposed in \cite{opt_isac}. To address this problem, a semi-definite relaxation (SDR) approach was employed. In \cite{ga_optim}, a Genetic Algorithm was utilized to address the resource allocation problem in full-duplex ISAC scenario.
Although the aforementioned works can achieve near-optimal solutions by approximating non-convex problems as convex ones or using the stochastic global optimization techniques, they require iterative optimization procedures that are computationally intensive. In dynamic environments with moving targets, where the channel can change rapidly, these classical approaches may struggle to adapt effectively in real-time as their complexity increases.

Recently, the rapid advancements in Artificial Intelligence (AI) and Machine Learning (ML), particularly in Deep Learning (DL) and Deep Reinforcement Learning (DRL), have introduced new approaches to solving traditional problems in communication and sensing systems. In \cite{dl_miso1} and \cite{dl_miso2}, DL-based beamforming methods are used to optimize beamforming vectors in downlink multi-user multiple-input single-output (MU-MISO) systems. However, these supervised learning approaches require a large amount of labeled training data. To address this challenge, Deep Reinforcement Learning (DRL), which combines DL and reinforcement learning (RL), offers a potential solution for real-time decision-making, as it can learn the environment through interaction and feedback. The Deep Deterministic Policy Gradient (DDPG) \cite{ddpg} is an off-policy DRL method that leverages a replay buffer to learn from past experiences and generates continuous actions, providing greater flexibility compared to discrete actions.
\subsection{Motivations and Contributions}
Building on the discussion above, our main contributions are as follows: 1) We derive the SINR expression for communication and sensing as performance metric, along with a joint beamforming and power allocation framework and a closed-form expression for the receiver beamformer. 2) We propose a transmission frame structure and a DRL framework based on DDPG algorithm. 3) We demonstrate through simulations that the proposed DRL-based solution achieves approximately 80\% of the communication sum-rate of an SDR-based method while reducing runtime significantly, and obtains 30\% improvement of the sum-rate with comparable runtime compared to Deep Q-Network (DQN), making it suitable for real-time ISAC applications. Furthermore, we show that the proposed framework can flexibly adjust ISAC performance by varying the weighting parameters.

$\textbf{Notations}$: Matrices are represented by bold uppercase letters, vectors by bold lowercase letters, and scalars by regular font. The transpose and Hermitian transpose are denoted by $(.)^T$ and $(.)^H$, respectively. The identity matrix of size $N\times N$ is denoted as $\boldsymbol{I}_N$, and $\mathcal{CN}(0, \sigma^2)$ denotes the complex Gaussian distribution with zero mean and variance $\sigma^2$. $Tr{\{.\}}$ is the trace of the enclosed item and the imaginary unit is denoted as $i^2=-1$.

\section{System model}

\begin{figure}[t]
\centerline{\includegraphics[width=5 cm]{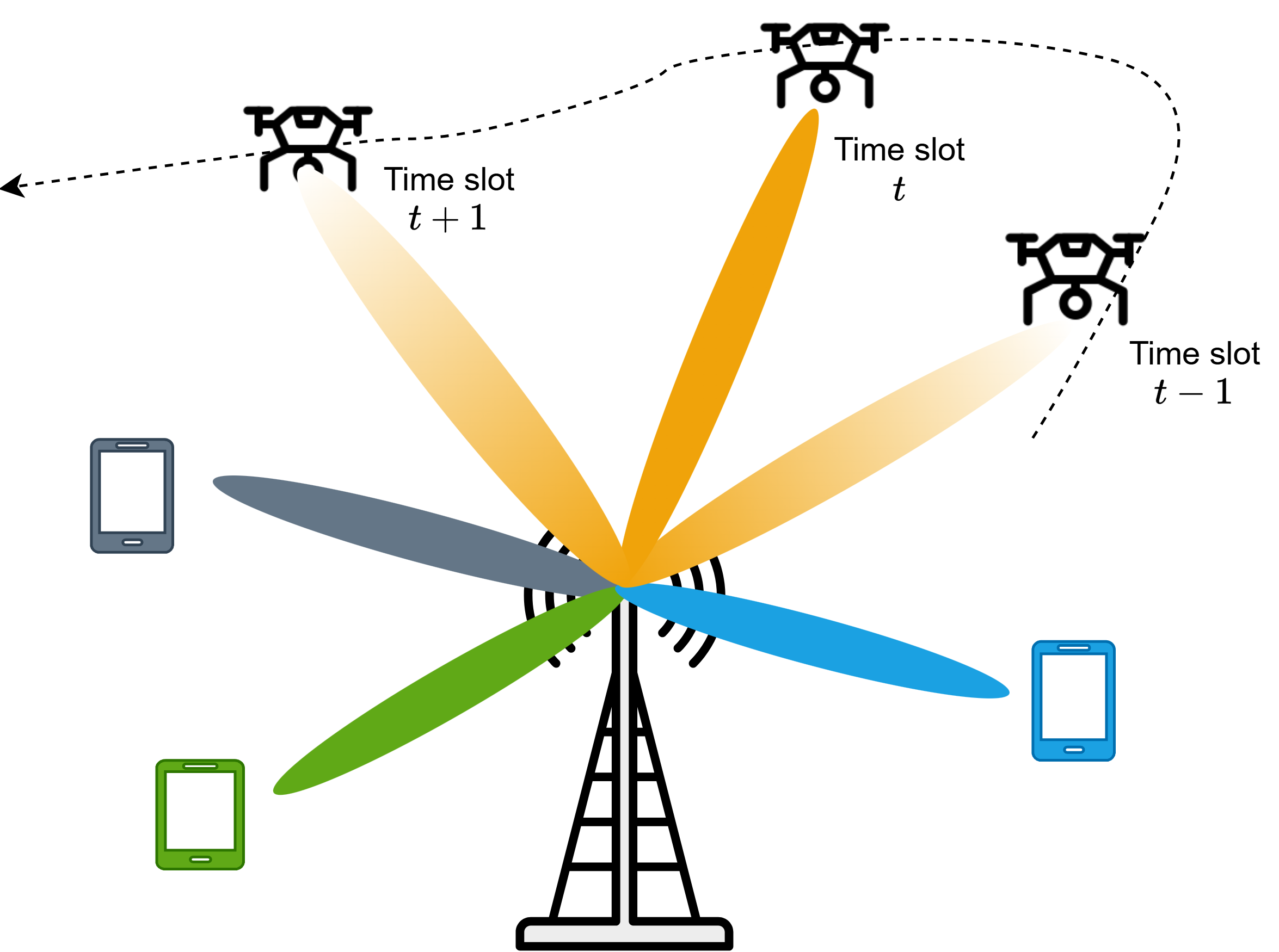}}
\caption{ISAC scenario. }\label{fig:system_config}
\end{figure}

We consider an ISAC system, as depicted in Fig. \ref{fig:system_config}, where the base station (BS) is equipped with two separate antenna arrays: a transmit array with $N_{\text{tx}}$ elements and a receive array with $N_{\text{rx}}$ elements. The BS and all users operate under a monostatic radar configuration, where the BS acts as both the transmitter and the receiver. In the transmit role, the BS simultaneously sends downlink signals to $J$ single-antenna users and performs target tracking by directing multiple beams toward both the users and the target. In the receive role, the BS collects echo signals reflected from the target. The transmit and receive arrays are placed in close proximity, ensuring that the target is observed from the same angles at both the transmitter and receiver.

\begin{figure}[t]
\centerline{\includegraphics[width=6.5 cm]{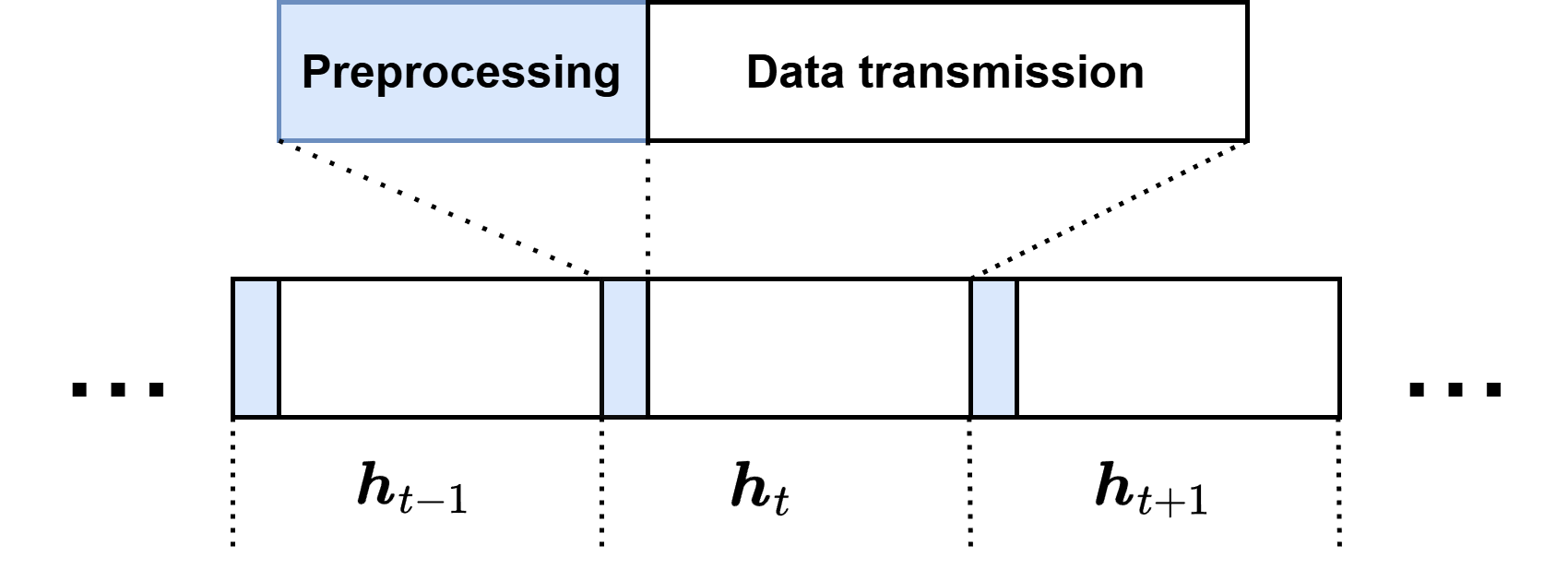}}
\caption{The proposed transmission frame structure for the dynamic beamforming problem.}\label{time_frame}
\end{figure}

The transmission frame structure is shown in Fig. \ref{time_frame}, where each frame is divided into two phases: the preprocessing phase and the transmission phase. We assume that channel state information (CSI) is available, thanks to advanced channel estimation techniques for ISAC systems \cite{DL_CSI}. In the first phase, the BS collects CSI and computes the beamformers based on a given beamforming strategy. In the subsequent phase, the BS serves both the users and targets using the computed beamforming matrices. In this work, we assume that the approximate position of the target is estimated during the initial beam scanning phase at the start of each time slot. This estimation is performed using a dedicated radar receiver integrated into the BS, which operates concurrently with the downlink synchronization process. During each time slot, dynamic beamforming is employed with the goal of maximizing the signal-to-interference-plus-noise ratio (SINR) toward the target. This SINR-maximizing approach can be further formulated as a problem that minimizes the target estimation error.

\subsection{Signal model}
First, we define the joint transmit signal at time slot $t$ as follows:
\begin{equation}\label{eq:transmit_signal}
    \boldsymbol{x}_t = \sum^J_{j=1}  \boldsymbol{w}_{j,t}^{c}  c^{c}_{j,t} + \boldsymbol{w}^{s}_t c^{s}_t,
\end{equation}
where $\boldsymbol{w}_{j,t}^c \in \mathbb{C}^{N_{\text{tx}}\times1}$ and $\boldsymbol{w}_{t}^s \in \mathbb{C}^{N_{\text{tx}}\times1}$ denote the beamforming vectors for communication user $j$ and the target at the time slot $t$, respectively. $c^{c}_{j,t}$ and $c^{s}_t$ are dedicated data symbols for user $j$ and target, respectively. We assume that they have unit power $\mathbb{E}\{c_{j,t}^cc_{j,t}^{c\space H}\}= 1$, $\mathbb{E}\{c_{t}^sc_{t}^{s\space H}\}= 1$, and are uncorrelated with each other.

Let $\boldsymbol{W}^c_t = [\boldsymbol{w}^c_{1,t},...\boldsymbol{w}^c_{J,t}] \in \mathbb{C}^{N_{\text{tx}}\times J}$ and $\boldsymbol{c}^c_t = [c_{1,t}^c,...c_{j,t}^c]^T \in \mathbb{C}^{J \times N_\text{tx}}$, Equation (\ref{eq:transmit_signal}) can be rewritten as:
\begin{equation}
    \boldsymbol{x}_t = \boldsymbol{W}_{t}^{c}  c^{c}_{t} + \boldsymbol{w}^{s}_t c^{s}_t.
\end{equation}
Let $\boldsymbol{b}(\theta, \phi) \in \mathbb{C}^{N \times 1}$ denotes the steering vector of an $N$-element antenna array for the given direction ($\theta, \phi$)
\begin{equation}
    \boldsymbol{b}(\theta,\phi) = \frac{1}{\sqrt{N}}[\exp(i\boldsymbol{k}(\theta,\phi)\boldsymbol{u}_1),...,\exp(i\boldsymbol{k}(\theta,\phi)\boldsymbol{u}_n)]^T,
\end{equation}
where $\boldsymbol{k}(\theta,\phi) = \frac{2\pi}{\lambda_c}[\sin(\theta)\cos(\phi), \sin(\theta)\sin(\phi),\cos(\theta)]^T \in \mathbb{C}^{N\times 1}$ describes the phase variation, $\lambda_c$ is the carrier wavelength, and $\boldsymbol{u}_n$ represents the position vector of the $n$-th antenna element. Hereafter, for illustration purposes, we denote $\boldsymbol{\Phi} = (\theta, \phi)$ as the set representing the corresponding 3D angular components (elevation and azimuth). 

\subsection{Communication model}
Assuming that the users are located within a maximum range of $d_\text{max}$ = 200 m, the maximum propagation time is $d_\text{max}/c_0 = 0.666 \mu s$, where $c_0$ is the speed of light. This duration is much smaller than the duration of the proposed time slot, which on the scale of a few milliseconds. Therefore, the signal's propagation delay can be disregarded in the formulation. The signal received by the user $j$ at the time slot $t$ is represented as
\begin{equation}
    y^{c}_{j,t} = \boldsymbol{h}_{j,t}^{c\space H}\boldsymbol{x}_t + n_{j,t},
\end{equation}
where $\boldsymbol{h}^c_{j,t} \in \mathbb{C}^{N_{\text{tx}} \times 1}$ is the channel between the BS and user $j$ at the time slot $t$, and $n_{j,t} \sim \mathcal{CN}(0, \sigma_n^2)$  denotes the additive white Gaussian noise (AWGN).

The communication channel consisting of $L$ propagation paths and can be expressed as:
\begin{equation}
\begin{aligned} 
  &  \boldsymbol{h}^c_{j,t} = \sqrt{N_{\text{tx}}}(\beta_{0,j,t} \boldsymbol{b}^{\text{tx}}_{j,t}(\boldsymbol{\Phi}_j) + \sum_{l=1}^{L-1} \beta_{l,j,t} \boldsymbol{b}^{\text{tx}}_{l,j,t}(\boldsymbol{\Phi}_l)),
  \end{aligned}
\end{equation}
where the first term and the second term is the LoS and NLoS path component, respectively. $\beta_{0,j,t}$ and $\beta_{l,j,t}$ are the complex fading coefficient of the LoS propagation path and of the $l$-th NLoS propagation path to user $j$ at the time slot $t$. $\boldsymbol{b}^{\text{tx}}_{j,t}(\boldsymbol{\Phi}_j) $ and $\boldsymbol{b}^{\text{tx}}_{l,j,t}(\boldsymbol{\Phi}_l)$ denote the transmit steering vector of the antenna array toward the user $j$ and for the $l$-th propagation path. 

The SINR of user $j$ at the time slot $t$ is 
\begin{equation}
    \nu^c_{j,t}=\frac{\left|\boldsymbol{h}_{j,t}^{c\space H} \boldsymbol{w}^c_{j,t}\right|^2}{\sum_{j^{\prime}=1, j^{\prime} \neq j}^J|\boldsymbol{h}_{j,t}^{c\space H} \boldsymbol{w}^c_{j^{\prime},t}|^2+ |\boldsymbol{h}_{j,t}^{c\space H} \boldsymbol{w}^s_{t}|^2+\sigma_{j,t}^2}.
\end{equation}

\subsection{Sensing model}
Similar to the communication model, we can ignore the round-trip propagation delay of the radar waves. In the monostatic radar system, the echo signal received at the BS can be expressed as:
\begin{equation}
    \boldsymbol{y}_{0,t} = \alpha_{0,t}\boldsymbol{A}_t(\boldsymbol{\Phi}_0)\boldsymbol{x}_t + \boldsymbol{z}_{0,t} \in \mathbb{C}^{N_{\text{rx}} \times 1}, 
\end{equation}
where $\boldsymbol{A}_t(\boldsymbol{\Phi}_0) = \boldsymbol{b}_t^{\text{rx}}(\boldsymbol{\Phi}_0)\boldsymbol{b}^{\text{tx} \space H}_t(\boldsymbol{\Phi}_0) $ with $\boldsymbol{b}_t^{\text{tx}}(\boldsymbol{\Phi})$ and $\boldsymbol{b}_t^{\text{rx}}(\boldsymbol{\Phi})$ denote the transmit and receive steering vector of the BS toward $\boldsymbol{\Phi}$ at the time slot $t$. $\boldsymbol{z}_{0,t} \sim \mathcal{CN}(0, \sigma_z^2\boldsymbol{I}_{N_{\text{rx}}})$ is the AWGN vector. $\alpha_{0,t}$ is the combined sensing channel gain that includes path loss and the radar cross section (RCS) of the target at the time slot $t$:
\begin{equation}
    \alpha_{0,t} = \sqrt{N_{\text{tx}} N_{\text{rx}}}\sqrt{\sigma_{0,t}}\frac{\lambda_c }{(4 \pi)^{3/2}{d_{0,t}}^2}\exp\left(\frac{-i2\pi (2d_{0,t})}{\lambda_c}\right),
\end{equation}
where $\sigma_{0,t}$ is the RCS of the target and $2d_{0,t}$ is the distance between the target and the BS at the time slot $t$. 

For the received echo signal at the BS, we apply a receive beamformer $\boldsymbol{u}^{s}_{t}$ at time slot $t$ to recover the sensing signal. The SINR for sensing can then be expressed as:
\begin{equation}\label{eq:sensing_sinr}
\nu^s_{t}=  \frac{|\boldsymbol{u}_{t}^{s\space H} \alpha_0\boldsymbol{A}_t(\boldsymbol{\Phi}_0) \boldsymbol{u}^s_{t}|^2}{\boldsymbol{u}_{t}^{s\space H}(\sum_{j=1}^J \alpha_0\boldsymbol{A}(\boldsymbol{\Phi}_0)\boldsymbol{w}_{j,t}  +\sigma_{BS}^2 \boldsymbol{I}_{N_{\text{rx}}}) \boldsymbol{u}^s_{t}}. 
\end{equation}

\section{Problem formulation}
\subsection{Optimization problem}
We employ the spectral efficiency as the evaluation for communication performance. For sensing, we use SINR as the metrics, which can further be translated to estimation accuracy. Let $\Gamma_t$ represent the spectral efficiency of all $J$ users at time slot $t$, which can be expressed as:
\begin{equation}
    \Gamma_{t} = \sum_{j=1}^J log_2 (1 + \nu^c_{j,t}).
\end{equation}
We formulate the optimization problem as maximizing the communication and sensing performance together, which can be written as:
\begin{subequations}\label{eq:optimization}
    \begin{align}
    &\underset{\substack{\{\boldsymbol{w}^c_{t,j}\}, \{\boldsymbol{w}^s_{t}\} }}{\operatorname{maximize}}  \;\; \rho \Gamma_t + (1-\rho) \nu^s_{t}, \\
    \text { s.t \;\;} 
    & Tr{\{\boldsymbol{W}_t^c\boldsymbol{W}_t^{cT}\}} + Tr{\{\boldsymbol{w}_t^s\boldsymbol{w}_t^{sT}\} \leq P_{\text{max}}}  \\
    & \left\|\boldsymbol{w}^c_{t,j}\right\|^2 \leq P_{0 }, \forall j, \quad\left\|\boldsymbol{w}^s_{t}\right\|^2 \leq P_{0 },
    \end{align}
\end{subequations}
where $\rho$ is the weight for trade-off between communication and sensing performance. Constraint (\ref{eq:optimization}b) ensures the maximum transmit power, and (\ref{eq:optimization}c) limits the maximum power per antenna element.  

\subsection{Receive beamforming closed-form solution}
To solve the problem (\ref{eq:optimization}) more efficiently, we rewrite it by introducing a closed-form solution for the receive beamformer. As can be seen, the SINR formulation of sensing is in the form of General Rayleigh Quotient. According to \cite{matrix_compute}, we can derive a close-form solution of $\boldsymbol{u}^s_{t}$ as:
\begin{equation}\label{eq14}
    \boldsymbol{u}_{t}^{RQ} = (\sum_{j=1}^J \alpha_{0,t}\boldsymbol{A}_t(\boldsymbol{\Phi}_0)\boldsymbol{w}_{t,j}  +\sigma_{BS}^2 \boldsymbol{I}_{N_{\text{rx}}})^{-1} \boldsymbol{a}_t^{\text{rx}}(\boldsymbol{\Phi}_0),
\end{equation}
By substituting $\boldsymbol{u}_{t}^{RQ}$ into (\ref{eq:sensing_sinr}), we obtain:
\begin{equation}
\begin{aligned}
\nu_{\text{t}}^s = &\left|\alpha_{0,t}\right|^2 \boldsymbol{a}_t^{\text{tx}H}(\boldsymbol{\Phi}_0) \boldsymbol{w}^s_{t}\boldsymbol{w}^{s\space H}_{t} \boldsymbol{A}_t^H(\boldsymbol{\Phi}_0)\\
& \times (\sum_{j=1}^J \alpha_{0,t}\boldsymbol{A}_t(\boldsymbol{\Phi}_0)\boldsymbol{w}_{t,j}  +\sigma_{BS}^2 \boldsymbol{I}_{N_{\text{rx}}})^{-1} \boldsymbol{a}_t^{\text{rx}}(\boldsymbol{\Phi}_0). 
\end{aligned}
\end{equation}
By applying the obtained receive beamformer to the original optimization problem (\ref{eq:optimization}), we derive a more tractable formulation with fewer optimization variables.

\subsection{Markov Decision Process (MDP) formulation}
In this section, the optimization problem (\ref{eq:optimization}) is presented as a MDP, which is the process can address most Reinforcement Learning problems. 
\subsubsection{Step and episode}
In our formulation, each episode $t$ corresponds to a time slot in the system in which the environment is observed.  Within each episode, the agent performs multiple steps $k$, making sequential decisions to optimize performance based on the observed channel.
The basic elements of MDP are state, action and reward, which are defined as follows: 
\subsubsection{State space $\mathcal{S}$} The state space tells the characteristics of the environment. $s_t(k) \in \mathcal{S}$ denotes the current characteristic of the environment at the step $k$ episode $t$, which consists of the current channel information $\boldsymbol{H}_t(k)$ and the beamforming vectors at the previous step $\boldsymbol{W}_t(k-1)$ in the same episode. $\boldsymbol{H}_t(k)$ consists of channel information of all J users, steering vectors and combined channel gain of target at the time slot $k$, equivalent to the set ${[ \boldsymbol{h}_{1,t}(k),...\boldsymbol{h}_{J,t}(k), \boldsymbol{a}_t^{\text{tx}}(k), \alpha_{0,t}(k) ] }$ The state $s_t(k)$ can be expressed as
\begin{equation}
    s_t(k) = [\boldsymbol{H}_t(k), \boldsymbol{W}^c_{t-1}(k), \boldsymbol{w}^s_{t-1}(k)].
\end{equation}
It should be noted that the neural network cannot deal with the complex number, so the real part and the imaginary part will be separated as the independent inputs to the network. It is also important to mention that the state space should be normalized to increase the learning capability of the agent. The normalization used will be mean normalization.
\subsubsection{Action space $\mathcal{A}$} The action space is a set of choices that agent can take during the learning process. Taking an action $a_t(k)$ at step $k$ of episode $t$, the state of environment will transit from $s_t(k)$ to $s_{t}(k+1)$ and get the reward $r_t(k)$. The actions are beamforming matrices.
\begin{equation}
    a_t(k) = [\boldsymbol{W}_t(k), \boldsymbol{w}^s_{t}(k)].
\end{equation}
\subsubsection{Reward} In order to maximize the objective value in the given optimization problem, the instant reward function at the step $k$, episode $t$ can be modeled as:
\begin{equation}
    r_t(k) = \rho \Gamma_t(k) + (1-\rho) \nu^s_{t}(k).
\end{equation}
To ensure fairness in the reward function, we notice that the first term is in logarithmic scale. Therefore, we rescale the second term to the same scale, resulting in the following modified reward function:
\begin{equation}
    r_t^{\prime}(k) = \rho \Gamma_t(k) + (1-\rho)log_{10}( \nu^s_{t}(k)).
\end{equation}
\section{DRL-based beamforming and power allocation}
In this section, we present a DRL framework for solving the joint beamforming and power allocation problem in a dynamic ISAC environment. 

\begin{figure}[t]
\centerline{\includegraphics[width=8.5 cm]{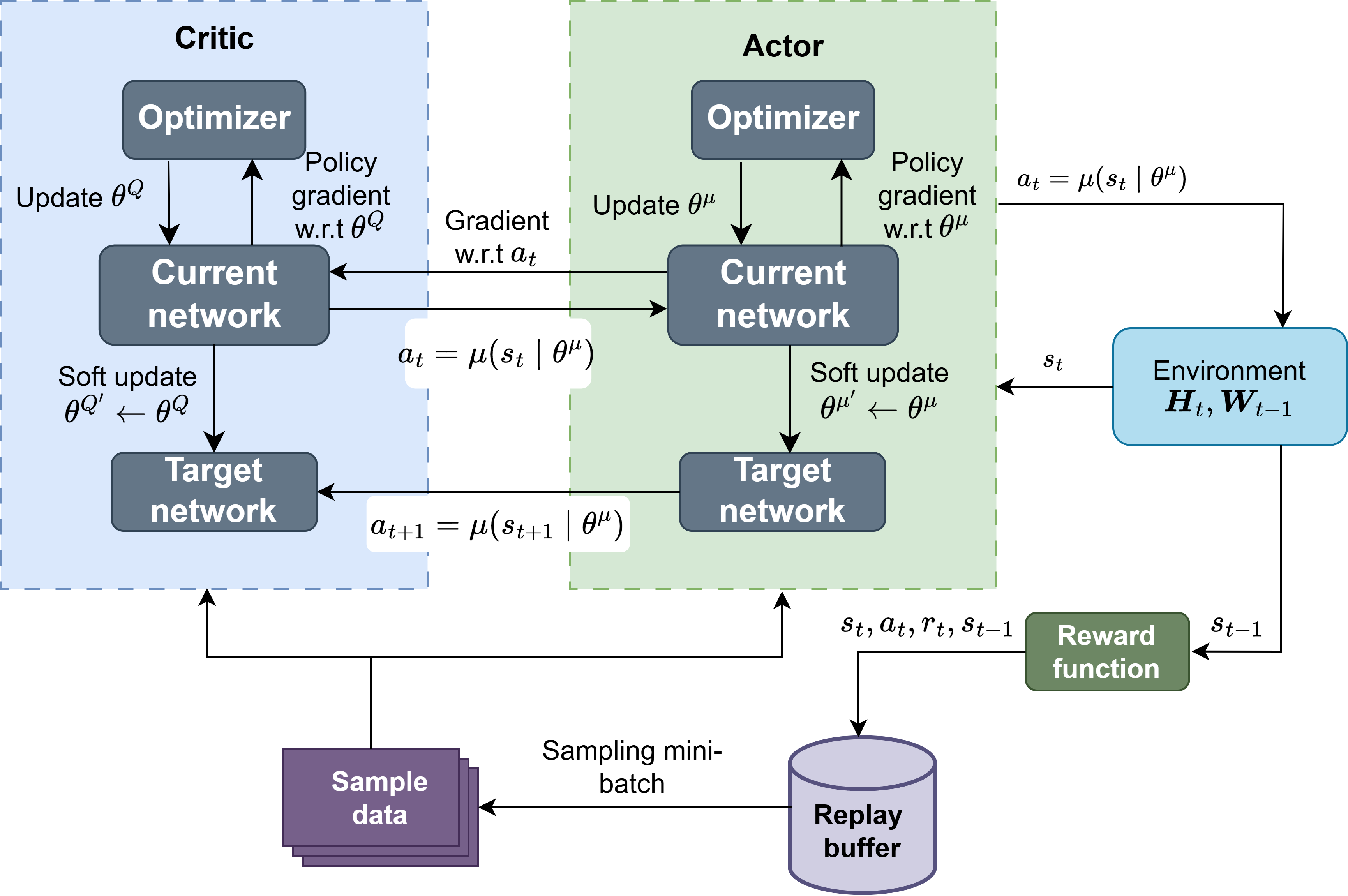}}
\caption{The structure of DDPG algorithm. }
\label{fig:ddpg}
\end{figure}
\subsection{Deep Deterministic Policy Gradient (DDPG) }
The Deep Deterministic Policy Gradient (DDPG) is an off-policy DRL method designed for continuous action spaces. It combines the advantages of deterministic policy gradients with Deep Neural Networks (DNNs), where the actor network outputs continuous actions and the critic network evaluates them through a Q-function. The policy gradient for the actor network is given as
\begin{equation}
    \nabla_{\vartheta^{\mu}} \mathcal{J}(\vartheta^\mu)=\nabla_{a_t(k)}Q(s_t(k),a_t(k)|\mu^Q)\nabla_{\vartheta^\mu}\mu(s_t(k)|\vartheta^{\mu})
\end{equation}
Here, $\mathcal{J}(\vartheta^\mu)$ is the expected return under the actor policy parameterized by $\vartheta^\mu$. The term $(\vartheta^\mu)=\nabla_{a_t(k)}Q(s_t(k),a_t(k)|\mu^Q)$ is the gradient of the critic (Q-function) with respect to the action, which measures how much the action affects the expected return. The second term, $\nabla_{\vartheta^\mu}\mu(s_t(k)|\vartheta^{\mu})$, is the gradient of the actor's output action with respect to its parameters.
The update training actor network function is expressed as:
\begin{equation}\label{eq:update_actor}
\vartheta^{\mu} \leftarrow \vartheta^{\mu} -  \mu_a\nabla_{\vartheta^{\mu}} \mathcal{J}(\vartheta^\mu), \end{equation}
where $\mu_a$ is the learning rate of the actor network.
To solve the proposed beamforming and power allocation problem, we explore the DDPG as its ability to handle the continuous state space and action space. The structure of DDPG is illustrated in Fig. \ref{fig:ddpg}, where there are two DNNs, the actor network and the critic network. The actor is updated by following the gradient of the expected return with respect to its parameters, which is estimated using the critic. Throughout learning process, they will update themselves together toward a better performance.

\subsubsection{Critic Network} The critic network evaluates the action $a_t(k)$ taken by the actor network in state $k$ episode $t$ $s_t(k)$ using the action-value function $Q(s_t(k),a_t(k)\mid\vartheta^Q)$, where $\vartheta^Q$ denotes the parameters of the critic network. It is trained by minimizing the temporal-difference (TD) error between the predicted $Q$-value and $\hat{Q}_t(k)$, which is the target $Q$-value and can be derived from the target networks:
\begin{equation} \label{critic_loss}\mathcal{L}(\vartheta^Q) = \left(Q(s_t(k), a_t(k) \mid \vartheta^Q) - \hat{Q}_t(k)\right)^2, \end{equation} 
\begin{equation} \hat{Q}_t(k) = r_t(k) + \gamma Q'\left(s_t(k+1), \mu'(s_t(k+1) \mid \vartheta^{\mu'}) \mid \vartheta^{Q'}\right), \end{equation}
where $\gamma$ is the discount rate.
The critic network is updated by gradient descent, and expressed as:
\begin{equation}\label{eq:update_critic}
\vartheta^{Q} \leftarrow \vartheta^{Q} -  \mu_c\nabla_{\vartheta^Q} \mathcal{L}(\vartheta^c), \end{equation}
where $\mu_Q$ is the learning rate of the critic network.
\subsubsection{Actor Network} At the step $k$ in episode $t$ with the state $s_t(k)$, the actor network selects an action $a_t(k)$ according to a deterministic policy $\mu$, parameterized by $\vartheta^{\mu}$. The actor network aims to maximize the expected return by learning the best mapping from states to actions:
\begin{equation}
    a_t(k) = \mu(s_t(k)\mid\vartheta^{\mu}).
\end{equation}
The gradient policy of the descent is computed based on the Q-value obtained by the critic network:
\begin{algorithm}[b]
\caption{Pseudocode of the proposed DDPG algorithm}
\begin{algorithmic} \label{alg:ddpg}
\State \textbf{Input:} $\boldsymbol{H}_t$
\State \textbf{Output:} optimal action $a_t = {\boldsymbol{W}_t, \boldsymbol{w}_t^s}$
\State \textbf{Initialization:} Replay buffer with memory size $\mathcal{D}$, training actor network, target actor network, training critic network, target critic network, initial beamforming matrices ${\boldsymbol{W}_t, \boldsymbol{w}_t^s}$.
\For{each episode t}
    \State Collect $\boldsymbol{H}_t$, $\boldsymbol{W}^c_{t-1}$, $\boldsymbol{w}^s_{t-1}$ for the $t^{th}$ episode
    \For{each step k}
    \State Obtain action $a_t(k)$, reward $r_t(k)$, next state $s_t(k+1)$ and store the experience in the replay memory
    \State Obtain Q-value from the critic network 
    \State Sample random batches from the replay memory
    \State Calculate critic loss (\ref{critic_loss})
    \State Update the critic training network (\ref{eq:update_critic})
    \State Update the training actor network (\ref{eq:update_actor})
    \State Every $U$ steps, update the target critic network (\ref{eq:update_critic_tgt})
    \State Every $U$ steps, update the target actor network (\ref{eq:update_actor_tgt})   
    \EndFor
\EndFor
\end{algorithmic}
\end{algorithm}
\subsubsection{Target Critic Network}
Similarly, a target critic network is maintained to produce a stable estimation of the Q-value during learning. It shares the same architecture as the main critic network and is updated in tandem with the target actor network. The target critic estimates $Q'(s, a \mid \vartheta^{Q'})$ and helps form the TD target for the critic loss:
\begin{equation}\label{eq:update_critic_tgt}
\vartheta^{Q'} \leftarrow \tau \vartheta^{Q} + (1 - \tau) \vartheta^{Q'}. \end{equation}
\subsubsection{Target Actor Network}
To stabilize training and prevent divergence caused by rapidly changing target values, DDPG introduces a separate target actor network. This network has the same structure as the main actor network but is updated more slowly. It provides a stable policy $\mu'(s \mid \vartheta^{\mu'})$ used when computing the target $Q$-value in the critic’s loss function. The parameters $\vartheta^{\mu'}$ are updated using a soft update mechanism:
\begin{equation}\label{eq:update_actor_tgt}
\vartheta^{\mu'} \leftarrow \tau \vartheta^{\mu} + (1 - \tau) \vartheta^{\mu'}, \end{equation}
Noting that, $\boldsymbol{W}_t^c$ and $\boldsymbol{w}_t^s$ have to satisfy the power constraint in \ref{eq:optimization}. To implement this, we introduce a normalization layer at the output of the actor network, in which $Tr{\{\boldsymbol{{W}_t^c}\boldsymbol{W}_t^{cT}\}} + Tr{\{\boldsymbol{w}_t^s\boldsymbol{w}_t^{sT}\}}=P_{\text{max}} $. The details of the proposed method are shown in Algorithm 1.
\section{Simulation results and analysis}
\subsection{Benchmarks}
To evaluate the performance of the proposed DDPG-based beamforming strategy, we compare it with several benchmark schemes:
\begin{itemize}
    \item SDR-based approach : The beamforming problem can be solved by using the method in \cite{sca} with some necessary modifications to fit our system model.  
    \item Deep Q-learning: The beamforming and power allocation strategies are adapted from \cite{DQN}, with modifications to fit our system model. In this benchmark, we focus on discrete beamforming in order to highlight the performance gains of continuous beamforming. Specifically, a beamforming codebook of size 512 is employed, and the transmit power is quantized into 10 discrete levels.  

    \item Random-based approach: The agent randomly chooses an action $a_t$ for all time slots, without any optimization. 
\end{itemize}
\subsection{Numerical parameters}
\begin{figure}[t]
\centerline{\includegraphics[width=8 cm]{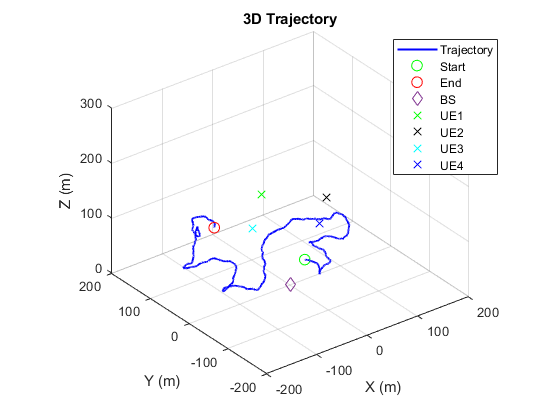}}
\caption{Visualization of the simulated scenario.}
\label{fig:trajectory}
\end{figure}

\begin{table}[t]
\caption{Numerical parameters}
\begin{center}
\begin{tabular}{ccc}
\hline
\textbf{Parameter} & \textbf{Notation} & \textbf{Value} \\
\hline
Operating frequency & $f_c$ & 39 GHz \\
Number of users & $J$ & 4  \\
Distance users to BS & $d_{j}$ & 150 m $\forall j$  \\
Distance target to BS & $d_{0}$ & 10 $\sim$ 150 m  \\
Total power & $P_\text{max}$ & 30 dBm \\
Time slot interval & $\Delta t$ & 20 ms \\
Noise power & n & -103 $dBm$ \\
ISAC weighting parameter & $\rho$ & 0.2 \\
Discount rate & $\gamma$ & 0.5 \\
Learning rate rate of actor and critic & $\mu_a$ = $\mu_c$ & $10^{-5}$ \\
Soft update rate of network & $\tau$ & $10^{-6}$ \\
Decay rate & $\lambda$ & $10^{-5}$ \\
Replay buffer size & $D$ & $10^{5}$ \\
Mini-batch size & $\mathcal{W}$ & 32 \\
Number of steps & $K$ & 20 \\
Number of episodes & $T$ & 5000 \\
Target network update interval &U & 2\\
\hline
\end{tabular}
\label{tab:parameters}
\end{center}
\end{table}
We select a uniform circular array (UCA) antenna with $N_{\text{tx}} = N_{\text{rx}} = 16$ elements, spaced half a wavelength apart, for both the transmitter and receiver at the BS.
The UCA has been shown to offer better spatial resolution, narrower beams, and deeper nulls compared to a rectangular array with the same number of elements \cite{uca}. 
We assume $L=10$ propagation paths, with the clusters uniformly distributed across the simulation environment. The locations of the target, users, and clusters are shown in Fig.~\ref{fig:trajectory}, where the target is assumed to follow the trajectory as in the figure. The complex path fading coefficient of the communication channel follows the model in \cite{channel_survey}. For simplicity, we assume an RCS of the target $\sigma_0 = 0$ $\space \text{dBm}^2$. The duration of time slot is set as 20 ms. We use the ReLU activation function for all layers of the DNN, except for the output layer of the actor network, which uses the tanh function. The detailed simulation parameters are provided in Table~\ref{tab:parameters}.

\begin{figure}[t]
\centerline{\includegraphics[width=8 cm]{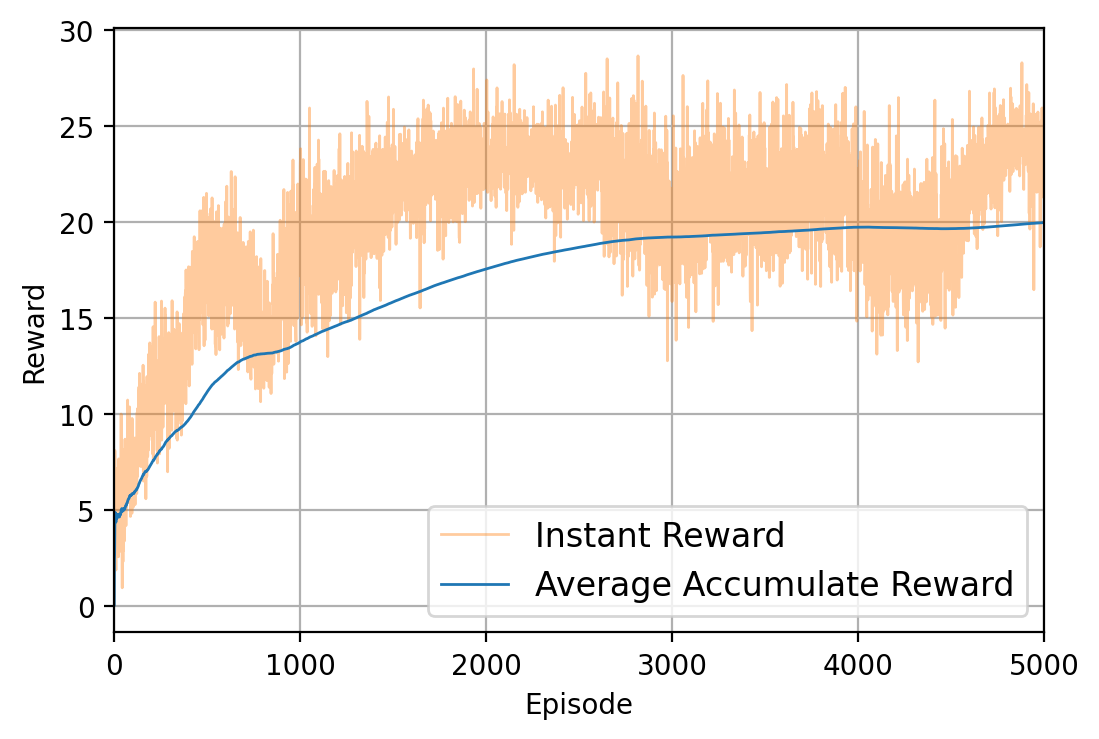}}
\caption{Instant and average accumulate reward over training episodes. }
\label{fig:convergence}
\end{figure}

\begin{figure}[t]
\centerline{\includegraphics[width=8 cm]{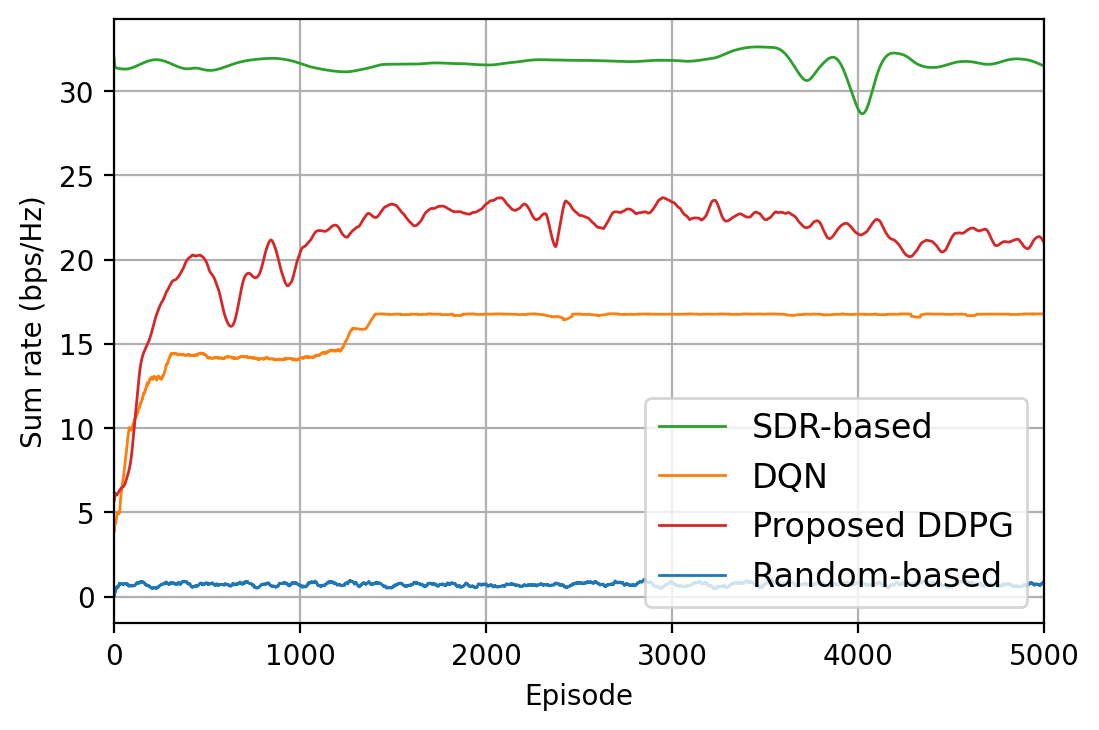}}
\caption{Average communication capacity of different approaches. Each value is a moving average of recent 50 episodes.}
\label{fig:capacity}
\end{figure}

\begin{figure}[t] 
    \subfigure[CDF of communication sum-rate]{\includegraphics[width=4.2 cm]{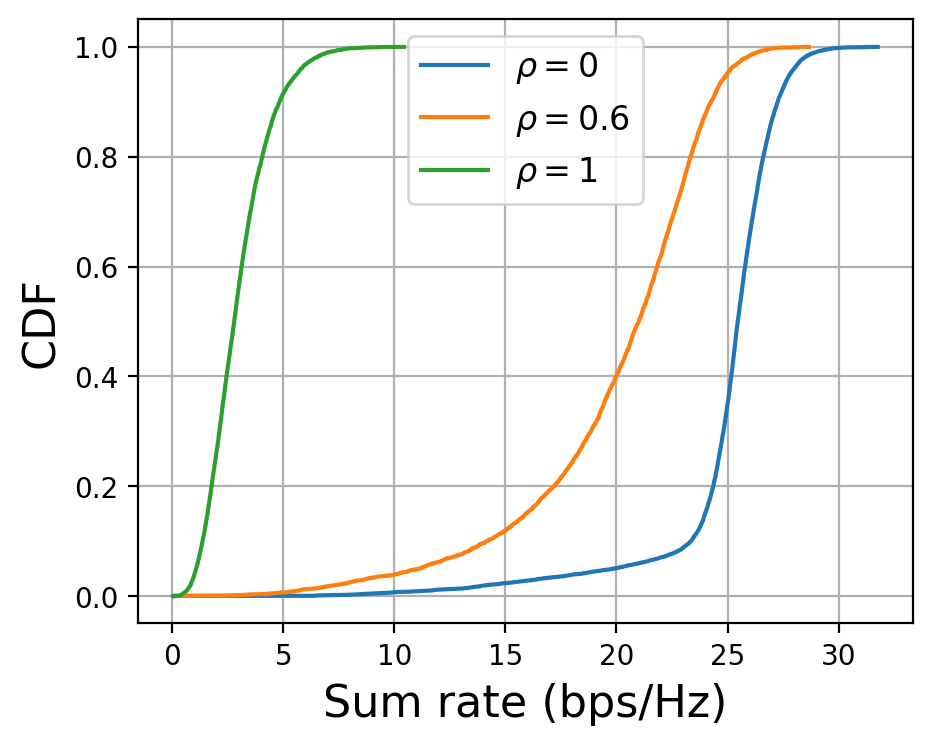}}
    \hfil
    \subfigure[CDF of sensing SINR]{\includegraphics[width=4.2 cm]{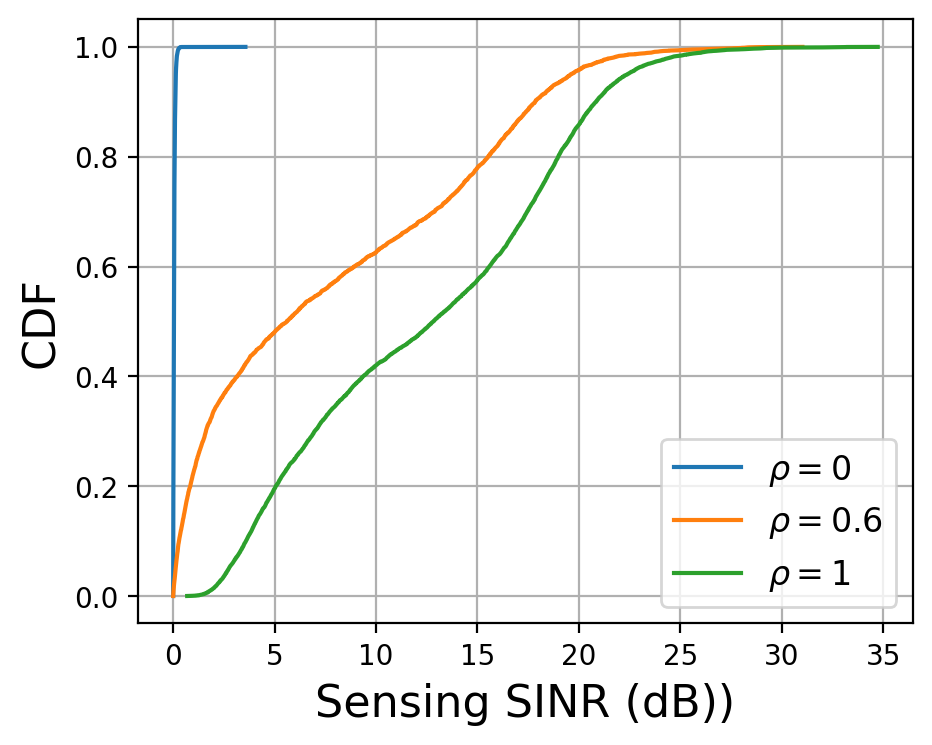}}
    \caption{CDF of varying ISAC weighting parameters.}
  
    \label{cdf}
\end{figure}
Fig.~\ref{fig:convergence} illustrates the learning process of the DDPG-based method, displaying both the instantaneous reward and the average reward over episodes. The average reward is computed as the moving average from the beginning to the current episode. It can be observed that the instantaneous reward fluctuates significantly during training; however, its general trend shows a gradual increase, indicating that the beamforming policy is being optimized. After approximately 2000 episodes, which corresponds to 400,000 ms, the learning process converges.

As shown in Fig. \ref{fig:capacity}, the SDR method achieves the highest communication capacity, while the random-based policy delivers the worst performance, as expected. The average communication capacity of the proposed DDPG method gradually improves throughout the training process, reaching approximately $80\%$ of the SDR performance.
Additionally, we compare the execution time required to obtain the solution in order to evaluate the computational complexity of the algorithm. Our simulation is run on an Intel Xeon 5220R CPU and an Nvidia A30 GPU. The average time for the agent to derive a decision is around 20 ms for 1 episode, whereas the SDR-based approach takes about 4500 ms, which is significantly higher than our proposed method. Therefore, in environments with fast dynamic changes, the SDR-based approach may be infeasible due to the large delay.
Compared with the DQN approach, our proposed algorithm achieves a higher performance in communication sum-rate while maintaining a comparable run time ($20$ ms for DDPG versus $17$ ms for DQN). This improvement is from the limitation of the DQN method, which is constrained by discrete beamforming and quantized power allocation, whereas our algorithm can exploit continuous optimization.  

Fig. \ref{cdf} demonstrates how the proposed DDPG algorithm manages the trade-off between communication and sensing performance through the ISAC weighting parameters $\rho$. When $\rho = 0$, the agent optimizes only for communication, achieving the highest sum-rate with most realizations above $25$ bps/Hz, but sensing performance falls near $0$ dB. Conversely, when $\rho = 1$, the algorithm prioritizes sensing, making sensing SINR above $10$ dB most cases, but the communication rate drops below $5$ bps/Hz. At $ \rho = 0.6$, the system achieves a balanced compromise, showing the agent's ability to adapt the resource allocation between two objectives. This analysis validates that the DRL framework can flexibly adjust ISAC performance depending on application requirements.

\section{Conclusion}
In this paper, we proposed a DRL-based framework for joint beamforming and power allocation in ISAC systems. The framework is designed to operate under dynamic conditions, where both the target and wireless channels vary over time. By utilizing the DDPG algorithm, we addressed the challenges of large-dimensional, time-varying optimization with a computationally efficient learning-based approach. The proposed method achieves convergence within 2000 training episodes and demonstrates the potential for real-time application, reducing computation time to approximately 20 ms compared to 4500 ms for the SDR-based baseline. While the communication sum rate reach up to 80\% of the SDR solution, the significant reduction in computation time makes the proposed method highly practical. Compared to the DQN benchmark with discrete beamforming, the proposed DDPG approach improves performance by around 30\% with comparable computational complexity. These results confirm that DRL-based solutions can effectively balance sensing and communication requirements under dynamic conditions, making them promising candidates for future ISAC systems. Future work will extend to wideband systems and target estimation and prediction.

\section*{Acknowledgment}
The authors appreciate the funding from Dutch Research Council (NWO) under the Open Technology Program (OTP) 3D-ComS (Project Nr. 19751).

\vspace{12pt}


\begin{thebibliography}{00}
\bibitem{isac_survey} R. Singh, A. Kaushik, M. Dajer, and M. Di Renzo, “ISAC standardization and synergies with key technology enablers: overview and future prospects,” Academic Press, 2025, pp. 397–408.

\bibitem{EE_optimization} Z. He, W. Xu, H. Shen, Y. Huang and H. Xiao, "Energy Efficient Beamforming Optimization for Integrated Sensing and Communication," in IEEE Wireless Commun. Lett. , vol. 11, no. 7, pp. 1374-1378, July 2022.
\bibitem{opt_isac} H. Hua, J. Xu, and Tony Xiao Han, “Optimal Transmit Beamforming for Integrated Sensing and Communication,” IEEE Trans. Veh. Technol., vol. 72, no. 8, pp. 10588–10603, Aug. 2023.

\bibitem{dl_miso1} J. Kim, H. Lee, S. -E. Hong and S. -H. Park, "Deep Learning Methods for Universal MISO Beamforming," in IEEE Wireless Commun. Lett. , vol. 9, no. 11, pp. 1894-1898, Nov. 2020
\bibitem{dl_miso2} W. Xia, G. Zheng, Y. Zhu, J. Zhang, J. Wang, and A. P. Petropulu, “A deep learning framework for optimization of MISO downlink
beamforming,” IEEE Trans. Commun., vol. 68, no. 3, pp. 1866–1880,
Mar. 2020.
\bibitem{ga_optim} D. N. Dao, H. Zhang, A. B. J. Kokkeler and Y. Miao, "Joint Beamforming for Multi-user Multi-target FD ISAC System: A Hybrid GRQ-GA Approach," 2025 IEEE Wireless Communications and Networking Conference (WCNC), Milan, Italy, 2025.
\bibitem{ddpg} Ebrahim Hamid Sumiea et al., “Deep deterministic policy gradient algorithm: A systematic review,” Heliyon, vol. 10, no. 9, pp. e30697–e30697, May 2024.
\bibitem{DL_CSI} H. He, C.-K. Wen, S. Jin, and G. Y. Li, “Deep Learning-Based Channel Estimation for Beamspace mmWave Massive MIMO Systems,” IEEE Wireless Communications Letters, vol. 7, no. 5, pp. 852–855, Oct. 2018.
\bibitem{matrix_compute} G. H. Golub and C. F. Van, Matrix computations. Baltimore: The Johns Hopkins University Press, Cop, 2013.
\bibitem{sca} Z. He, W. Xu, H. Shen, D. Wing, Y. C. Eldar, and X. You, “Full-Duplex Communication for ISAC: Joint Beamforming and Power Optimization,” IEEE Journal on Selected Areas in Communications, vol. 41, no. 9, pp. 2920–2936, Sep. 2023.
\bibitem{DQN} F. B. Mismar, B. L. Evans and A. Alkhateeb, "Deep Reinforcement Learning for 5G Networks: Joint Beamforming, Power Control, and Interference Coordination," in IEEE Trans. Commun., vol. 68, no. 3, pp. 1581-1592, March 2020.
\bibitem{uca} P. Ioannides and C. A. Balanis, "Uniform circular and rectangular arrays for adaptive beamforming applications," in Antennas Wireless Propag. Lett, vol. 4, pp. 351-354, 2005.
\bibitem{channel_survey} Z. Wei et al., “Integrated Sensing and Communication Channel Modeling: A Survey,” IEEE Internet of Things Journal, pp. 1–1, Jan. 2024.
\end{thebibliography}
\end{document}